\documentclass{stacs_proc}

\usepackage{graphicx,natbib,amssymb}
\usepackage{moreverb}

\theoremstyle{plain}

\begin{document}

\title[Jordan Curve Theorem in Coq with hypermaps]{Discrete Jordan
Curve Theorem:\\ A proof formalized in Coq with hypermaps}

\author[lab1]{J.-F. Dufourd}{Jean-Fran\c{c}ois Dufourd}
\address[lab1]{Universit\'{e} Louis-Pasteur de Strasbourg,
UFR de Math\'{e}matique et d'Informatique,
\newline
Labo. des Sciences de l'Image, de l'Informatique et de la T\'{e}l\'{e}d\'{e}tection
(UMR CNRS-ULP 7005),\newline
P\^{o}le API,
Boulevard S\'{e}bastien Brant, 67400 Illkirch, France}  
\email{dufourd@dpt-info.u-strasbg.fr}  
\thanks{Acknowledgements: This research is supported by the "white" project GALAPAGOS, French ANR, 2007} 


\begin{abstract}
  \noindent This paper presents a formalized proof of a discrete form of the Jordan Curve Theorem. It is based on a hypermap model of planar subdivisions, formal specifications and proofs assisted by the Coq system. Fundamental properties are proven by structural or noetherian induction: Genus Theorem, Euler's Formula, constructive planarity criteria. A notion of ring of faces is inductively defined and a Jordan Curve Theorem is stated and proven for any planar hypermap.
\end{abstract}

\maketitle

\stacsheading{2008}{253-264}{Bordeaux}
\firstpageno{253}

\section*{Introduction}\label{S:one}
This paper presents a formal statement and an assisted proof of a Jordan Curve Theorem (JCT) discrete version. In its common form, the theorem says that the complement of a continuous simple closed curve (a Jordan curve) $C$ in an affine real plane is made of two connected components whose border is $C$, one being bounded and the other not. The discrete form of JCT we deal with states that in a finite subdivision of the plane, breaking a ring $R$ of faces increases by $1$ the connectivity of the subdivision. It is a weakened version of the original theorem where the question of bound is missing. 
However, it is widely used in computational geometry and discrete geometry for imaging, where connection is the essential information \cite{ros79, fra95}. In fact, we only 
are in a combinatoric framework, where any embedding is excluded, and where bounding does not make sense.

In computational topology, subdivisions are best described by map models, the most general being {\em hypermaps} \cite{tut1, cor}. We propose a purely combinatorial proof of JCT based on this structure. The hypermap framework is entirely formalized and the proofs are developed interactively and verified by the Coq proof assistant \cite{Coq}. Using an original way to model, build and destruct hypermaps, the present work brings new simple constructive planarity and connectivity criteria. It proposes a new direct expression of JCT and a simple constructive proof with algorithmic extensions. 
It is also a large benchmark for the software specification framework we have been developing in the last fifteen years for map models used in geometric modeling and computer imagery \cite{ber:duf, duf07a, duf07b}.  

The useful Coq features are reminded and the whole process is described, but the full details of the proofs are omitted. Section \ref{RW} summarizes related work. 
Section \ref{MA} recalls some mathematical materials.
Section \ref{FH} proposes basic hypermap specifications.
Section \ref{PC} proves constructive criteria of hypermap planarity and connectivity.
Section \ref{FR} inductively specifies the rings and their properties. 
Section \ref{JCT} proves the discrete JCT. 
Section \ref{Cl} concludes. 

\section{Related work}
\label{RW}

The JCT is a result of classical plane topology, first stated by C. Jordan in 1887, but of which O. Veblen gives the first correct proof in 1905. In 1979, W.T. Tutte proposes operations and properties of combinatorial maps, {\em e.g.} planarity and Euler's Formula, defines rings and proves a discrete JCT \cite{tut1}. Our theorem statement is comparable, but our framework is modeled differently and all our proofs are formalized and computer-assisted. 

In 2003, G. Bauer and T. Nipkow specify planar graphs and triangulations in Isabelle/Isar to carry out interactive proofs of Euler's Formula and of the Five Colour Theorem \cite{bau:nip}. However, they do not approach the JCT. In 2005, A. Kornilowicz designs for the MIZAR project a semi-automated classical proof of a continuous form of JCT in an Euclidean space \cite{kor}. In 2005 also, on his way towards the proof of the Kepler conjecture in the Flyspeck projet, T. Hales proves the JCT for planar rectangular grids with the HOL Light system, following the Kuratowski characterization of planarity \cite{hal}. 

In 2005 always, G. Gonthier {\em et al.} prove the Four Colour Theorem using Coq. Plane subdivisions are described by hypermaps, and Euler's Formula is used as a global planarity criterion \cite{gon}. A local criterion, called {\em hypermap Jordan property}, is proven equivalent. The main part of this work is the gigantic proof of the Four Colour Theorem with hypermaps and sophisticated proof techniques. The hypermap formalization is very different from ours and it seems that JCT is not explicitly proven there. Finally, since 1999, we carry out experiments with Coq for combinatorial map models of space subdivisions \cite{deh:duf2,duf07a,duf07b}.

\section{Mathematical Aspects}
\label{MA}
\begin{figure}
\begin{center}
\includegraphics*[scale =.55]{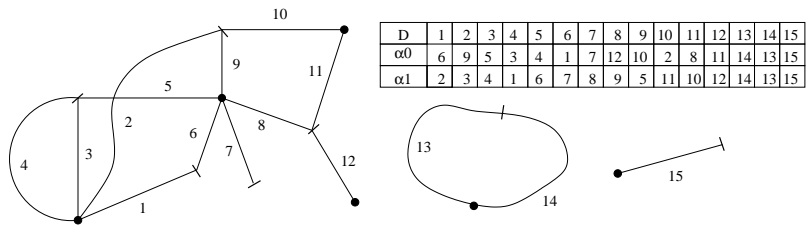}
\end{center}
\caption{An example of hypermap.}
\label{fig:Uhmap1}
\end{figure}

\begin{defi}[Hypermap]
\label{HD}
A {\em hypermap} is an algebraic structure $M = (D, \alpha_0,
\alpha_1)$, where $D$ is a finite set whose elements are
called {\em darts}, and $\alpha_0$, $\alpha_1$ are permutations on
$D$.

If $y= \alpha_k(x)$, $y$ is the $k$-{\em successor} of $x$, $x$ is
the $k$-{\em predecessor} of $y$, and $x$ and $y$ are said to be
$k$-{\em linked}.
\end{defi}
In Fig. \ref{fig:Uhmap1}, as functions  $\alpha_0$ and  $\alpha_1$ on $D = \{1,\ldots, 15\}$ are permutations, $M = (D, \alpha_0, \alpha_1)$ is a hypermap. It is drawn on the plane by associating to each dart a curved arc oriented from a bullet to a small stroke: $0$-linked (resp. $1$-linked) darts share the same small stroke (resp. bullet). By convention, in the drawings of hypermaps on surfaces, $k$-successors turn counterclockwise around strokes and bullets. Let $M = (D, \alpha_0, \alpha_1)$ be a hypermap. 

\begin{defi}(Orbits and hypermap cells)\label{OC}\\
(1) Let $f_1,\dots,f_n$ be $n$ functions in $D$. The {\em orbit} of $x \in D$ for $f_1,\dots,f_n$ is the subset of $D$ denoted by $\langle f_1,\dots,f_n \rangle(x)$, the elements of which are accessible from $x$ by any composition of $f_1,\dots,f_n$.\\
(2) In $M$,
$\langle \alpha_0 \rangle (x)$ is the $0$-orbit or {\em edge} of dart $x$, $\langle \alpha_1 \rangle(x)$ its $1$-orbit or {\em vertex}, $\langle \phi \rangle(x)$ its {\em face} for $\phi=\alpha_1^{-1} \circ \alpha_0^{-1}$, and $\langle \alpha_0, \alpha_1\rangle(x)$ its {\em (connected) component}.
\end{defi}
In Fig. \ref{fig:Uhmap1} the hypermap contains $7$ edges (strokes), $6$ vertices (bullets), $6$ faces and $3$ components. 
For instance, $\langle \alpha_0 \rangle(3)=\{3,5,4\}$ is the edge of dart $3$, $\langle \alpha_1 \rangle(3)=\{3,4,1,2\}$ its vertex. 
Faces are defined, through $\phi$, for a dart traversal in counterclockwise order, when the hypermap is drawn on a surface. Then, every face which encloses a bounded (resp. unbounded) region on its left is called {\em internal} (resp. {\em external}). In Fig. \ref{fig:Uhmap1}, the (internal) face of $8$ is $\langle \phi \rangle(8)= \{8,10\}$ and the (external) face of $13$ is $\langle \phi \rangle (13)= \{13\}$.
Let $d,e,v,f$ and $c$ be the numbers of darts, edges, vertices, faces and components of $M$. 

\begin{defi}(Euler characteristic, genus, planarity)\\
(1) The {\em Euler characteristic} of $M$ is $\chi= v + e + f - d$.\\
(2) The {\em genus} of $M$ is $g = c -\chi / 2$.\\
(3) When $g=0$, $M$ is said to be {\em planar}.
\end{defi}

For instance, in Fig. \ref{fig:Uhmap1}, $\chi= 6 + 6 + 7 - 15 = 4$ and $g = 3 - \chi / 2 = 1$. Consequently, the hypermap is non planar. These values satisfy the following results:

\begin{thm}[of the Genus]
$\chi$ is an even integer and $g$ is a natural number.
\end{thm}

\begin{cor}[Euler Formula]
A non empty connected $-$ i.e. with $c=1$ $-$ planar hypermap satisfies $v+e+f-d = 2$.
\end{cor}

When $D \neq \emptyset$, the {\em representation} of $M$ on an {\em orientable closed} surface is a mapping of edges and vertices onto points, darts onto open oriented Jordan arcs, and faces onto open connected regions. It is an {\em embedding} when every component of $M$ realizes a partition of the surface. 
Then, the genus of $M$ is the minimum number of {\em holes} in an orientable closed surface where such an embedding is possible, thus drawing a subdivision, or a polyhedron, by hypermap component \cite{gri}. 
For instance, all the components of the hypermap in Fig. \ref{fig:Uhmap1} can be embedded on a torus ($1$ hole) but not on a sphere or on a plane ($0$ hole). When a (planar) hypermap component is embedded on a plane, the corresponding subdivision has exactly one unbounded (external) face. But a non planar hypermap can never be embedded on a plane: in a drawing on a plane, some of its faces are neither internal nor external, {\em e.g.} $\langle \phi \rangle(1)= \{1,5,2,11,12,7,6,4,9\}$ in Fig. \ref{fig:Uhmap1}. Conversely, any subdivision of an {\em orientable closed} surface can be modeled by a hypermap. In fact, the formal presentation which follows is {\em purely combinatorial}, {\em i.e} without any topological or geometrical consideration. 

\vskip-.3cm
\subsection{Rings of faces and Jordan Curve Theorem}
To state the version of JCT we will prove, we need the concepts of {\em double-link}, {\em adjacent faces} and {\em ring of faces} in a hypermap $M=(D,\alpha_0, \alpha_1)$.

\begin{defi}(Double-link and adjacent faces)\\
(1) A {\em double-link} is a pair of darts $(y, y')$ where $y$ and $y'$ belong to the same edge.\\
(2) The faces $F$ and $F'$ of $M$ are said to be {\em adjacent by the double-link} $(y, y')$ if $y$ is a dart of $F$ and $y'$ a dart of $F'$.
\end{defi}

We choose a face adjacency {\em by an edge} rather than {\em by a vertex} as does W.T. Tutte \cite{tut1}. In fact, due to the homogeneity of dimensions $0$ and $1$ in a hypermap, both are equivalent. 

\begin{defi}(Ring of faces)
\label{RoF}\\
A {\em ring of faces} $R$ of length $n$ in $M$ is a non empty sequence of double-links $(y_i, y'_i)$, for $i=1,\ldots, n$, with the following properties, where $E_i$ and $F_i$ are the edge and face of $y_i$: \\
(0) {\em Unicity:} $E_i$ and $E_j$ are distinct, for $i,j=1,\ldots, n$ and $i \neq j$;\\
(1) {\em Continuity:} $F_i$ and $F_{i+1}$ are adjacent by the double-link $(y_i, y'_i)$, for $i=1,\ldots, n-1$;\\
(2) {\em Circularity}, or {\em closure:} $F_n$ and $F_1$ are adjacent by the double-link $(y_n, y'_n)$;\\
(3) {\em Simplicity:} $F_i$ and $F_j$ are distinct, for $i,j=1,\ldots, n$ and $i \neq j$.
\end{defi}

\begin{figure}
\begin{center}
\includegraphics*[scale =.55]{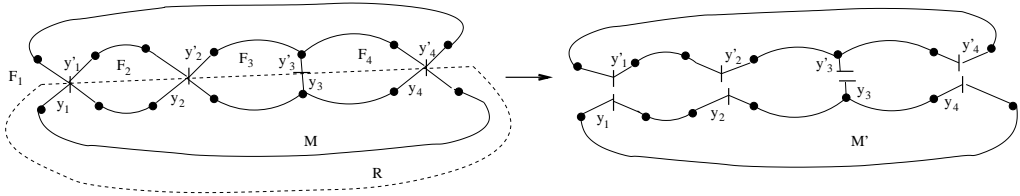}
\end{center}
\caption{Break of $M$ along a ring $R$ of length $n = 4$ giving $M'$.}
\label{fig:Uring4bis}
\end{figure}

This notion simulates a Jordan curve represented in dotted lines in Fig. \ref{fig:Uring4bis} on the left for $n = 4$. Then, we define the {\em break along a ring}, illustrated in Fig. \ref{fig:Uring4bis} on the right. 

\begin{defi}(Break along a ring)
\label{BaR}\\
Let $R$ be a ring of faces of length $n$ in $M$. Let $M_i=(D,\alpha_{0,i}, \alpha_1)$, for $i=0,\ldots, n$, be a hypermap sequence, where the $\alpha_{0,i}$ are recursively defined by:\\
(1) $i = 0 $: $\alpha_{0,0}$ = $\alpha_0$;\\
(2) $1 \leq i \leq n $: for each dart $z$ of $D$:
 $\alpha_{0,i}(z)$ = 
    if $\alpha_{0,i-1}(z) = y_i$ then $y'_i$ 
    else if $\alpha_{0,i-1}(z) = y'_i$ then $y_i$ else $\alpha_{0,i-1}(z).$ \\
Then, $M_n=(D,\alpha_{0,n}, \alpha_1)$ is said to be obtained from $M$ by a {\em break along} $R$.  
\end{defi}

Finally, the theorem we will prove in Coq mimics the behaviour of a cut along a simple Jordan curve of the plane (or of the sphere) into two components:

\begin{thm}[Discrete Jordan Curve Theorem]
Let $M$ be a planar hypermap with $c$ components, $R$ be a ring of faces in $M$, and $M'$ be the break of $M$ along $R$. The number $c'$ of components of $M'$ is such that $c'=c+1$.
\end{thm} 

\section{Hypermap specifications}
\label{FH}

\vskip-.3cm
\subsection{Preliminary specifications}
\label{PS}

In Coq, we first define an inductive type {\tt dim} for the two dimensions at stake:
\begin{verbatimtab}
  Inductive dim:Set:= zero: dim | one: dim.
\end{verbatimtab}
All objects being typed in Coq, {\tt dim}
has the type {\tt Set} of all concrete types. Its  {\em constructors} are the constants {\tt zero} and {\tt one}. In each inductive type, the generic equality predicate $=$ is built-in but its decidability is not, because Coq's logic is intuitionistic. For {\tt dim}, the latter can be established as the lemma:
\begin{verbatimtab}
  Lemma eq_dim_dec: forall i j : dim, {i=j}+{~i=j}.
\end{verbatimtab}
Once it is made, its proof is an object of the sum type {\tt \{i=j\}+\{}\verb\~\{\tt i=j\}}, {\em i.e.} a function, named {\tt eq}$\_${\tt dim}$\_${\tt dec}, that tests whenever its two arguments are equal. The lemma is interactively proven with some tactics, the reasoning being merely a structural induction on both {\tt i} and {\tt j}, here a simple case analysis. Indeed, from each inductive type definition, Coq generates an {\em induction principle}, usable either to prove propositions or to build total functions on the type. We identify the type {\tt dart} and its equality decidability {\tt eq}$\_${\tt dart}$\_${\tt dec} with the built-in {\tt nat} and {\tt eq}$\_${\tt nat}$\_${\tt dec}. Finally, to manage exceptions, a {\tt nil} dart is a renaming of {\tt 0}:
\begin{verbatimtab}
  Definition dart:= nat.
  Definition eq_dart_dec:= eq_nat_dec.
  Definition nil:= 0.
\end{verbatimtab}

\vskip-.3cm
\subsection{Free maps}
\label{FM}
\begin{figure}
\begin{center}
\includegraphics*[scale =.55]{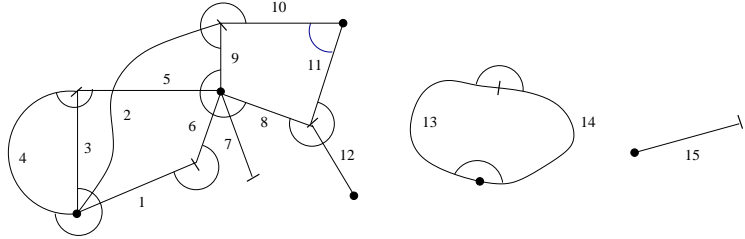}
\end{center}
\caption{A hypermap with its incompletely linked orbits.}
\label{fig:hmap2}
\end{figure}

The hypermaps are now approached by a general notion of {\em free map}, thanks to a free algebra of terms of inductive type {\tt fmap} with $3$ constructors, {\tt V}, {\tt I} and {\tt L}, respectively for the {\em empty} (or {\em void}) map, the {\em insertion} of a dart, and the {\em linking} of two darts:
\begin{verbatimtab}
  Inductive fmap:Set:=
     V : fmap | I : fmap->dart->fmap | L : fmap->dim->dart->dart->fmap.
\end{verbatimtab}
For instance, the hypermap in Fig. \ref{fig:Uhmap1} can be modeled by the free map represented in Fig.\ref{fig:hmap2} where the $0$- and $1$-links by {\tt L} are represented by arcs of circle, and where the orbits remain open. Again, Coq generates an induction principle on free maps. 

Next, {\em observers} of free maps can be defined. The predicate {\tt exd} express that a dart exists in a hypermap. Its definition is recursive, which is indicated by {\tt Fixpoint}, thanks to a pattern matching on {\tt m} written {\tt match m with...}. The attribute $\{${\tt struct m}$\}$ allows Coq to verify that the recursive calls are performed on smaller {\tt fmap} terms, thus ensuring termination. 
The result is {\tt False} or {\tt True}, basic constants of {\tt Prop}, the built-in type of propositions.
Note that terms are in prefix notation and that $\_$ is a place holder:

\begin{verbatimtab}
  Fixpoint exd(m:fmap)(z:dart){struct m}:Prop:=
   match m with
     V => False | I m0 x => z=x \/ exd m0 z | L m0 _ _ _  => exd m0 z
   end.
\end{verbatimtab}
The decidability {\tt exd}$\_${\tt dec} of {\tt exd} directly derives, thanks to a proof by induction on {\tt m}. Then, a version, denoted {\tt A}, of operation $\alpha_k$ of Definition \ref{HD} completed with {\tt nil} for convenience is written as follows, the inverse {\tt A}$\_${\tt 1} being similar:

\begin{verbatimtab}
  Fixpoint A(m:fmap)(k:dim)(z:dart){struct m}:dart:=
   match m with
     V => nil | I m0 x => A m0 k z | L m0 k0 x y => 
       if eq_dim_dec k k0 then if eq_dart_dec z x then y else A m0 k z 
       else A m0 k z
   end.
\end{verbatimtab}
Predicates {\tt succ} and {\tt pred} express that a dart has a {\tt k}-successor and a {\tt k}-predecessor (not {\tt nil}), with the decidabilities {\tt succ}$\_${\tt dec} and {\tt pred}$\_${\tt dec}. In hypermap {\tt m} of Fig. \ref{fig:hmap2}, {\tt A m zero 4 = 3, A m zero 5 = nil, succ m zero 4 = True, succ m zero 5  = False, A}$\_${\tt 1 m one 2 = 1}. In fact, when a {\tt k}-orbit remains open, which will be required in the following, we can obtain its {\tt top} and {\tt bottom} from one of its dart {\tt z}. Then, we can do as if the {\tt k}-orbit were closed, thanks to the operations {\tt cA} and {\tt cA}$\_${\tt 1} which {\em close} {\tt A} and {\tt A}$\_${\tt 1}, in a way similar to operation $K$ of W.T. Tutte \cite{tut1}. For instance, in Fig. \ref{fig:hmap2}, {\tt top m one 1 = 3, bottom m one 1 = 4, cA m one 3 = 4, cA}$\_${\tt 1 m one 4 = 3}. 

Finally, {\em destructors} are also recursively defined. First, {\tt D:fmap->dart->fmap} deletes the latest insertion of a dart by {\tt I}. Second, {\tt B, B}$\_${\tt:fmap->dim->dart->fmap} break the latest {\tt k-}link inserted for a dart by {\tt L}, forward and backward respectively. 

\vskip-.3cm
\subsection{Hypermaps}
\label{H}
Preconditions written as predicates are introduced for {\tt I} and {\tt L}:

\begin{verbatimtab}
  Definition prec_I(m:fmap)(x:dart):= x <> nil /\ ~ exd m x.
  Definition prec_L(m:fmap)(k:dim)(x y:dart):=
   exd m x /\ exd m y /\ ~ succ m k x /\ ~ pred m k y /\ cA m k x <> y.
\end{verbatimtab}
If {\tt I} and {\tt L} are used under these conditions, the free map built necessarily has open orbits. In fact, thanks to the closures {\tt cA} and {\tt cA}$\_${\tt 1}, it can always be considered as a true hypermap exactly equipped with operations $\alpha_k$ of Definition \ref{HD}. It satisfies the {\em invariant}:

\begin{verbatimtab}
  Fixpoint inv_hmap(m:fmap):Prop:=
   match m with
      V => True | I m0 x => inv_hmap m0 /\ prec_I m0 x
    | L m0 k0 x y => inv_hmap m0 /\ prec_L m0 k0 x y
   end.
\end{verbatimtab}
Such a hypermap was already drawn in Fig. \ref{fig:hmap2}. Fundamental proven properties are that, for any {\tt m} and {\tt k}, {\tt (A m k)} and {\tt (A}$\_${\tt 1 m k)} are {\em injections} inverse of each other, and {\tt (cA m k)} and {\tt (cA}$\_${\tt 1 m k)} are {\em permutations} inverse of each other, and are closures. 
Finally, traversals of faces are based on function {\tt F} and its closure {\tt cF}, which correspond to $\phi$ (Definition \ref{OC}).
So, in Fig. \ref{fig:hmap2}, {\tt F m 1 = nil, cF m 1 = 5}. Properties similar to the ones of {\tt A}, {\tt cA} are proven for {\tt F}, {\tt cF} and their inverses {\tt F}$\_${\tt 1}, {\tt cF}$\_${\tt 1}.

\vskip-.3cm
\subsection{Orbits}
\label{O}
Testing if there exists a path from a dart to another
in an orbit for a hypermap permutation is of prime importance, for instance to determine the number of orbits. The problem is exactly the same for $\alpha_0$, $\alpha_1$ or $\phi$ (Definitions \ref{HD} and \ref{OC}). That is why a {\em signature} {\tt Sigf} with formal parameters {\tt f}, {\tt f}$\_${\tt 1} and their properties is first defined. 

Next, a {\em generic module} (or {\em functor}) {\tt Mf(M:Sigf)}, the formal parameter {\tt M} being a {\em module} of type {\tt Sigf}, is written in Coq to {\em package} generic definitions and proven properties about {\tt f} and {\tt f}$\_${\tt 1}. Among them, we have that each {\tt f}-orbit of {\tt m} is {\em periodic} with a positive smallest uniform period for any dart {\tt z} of the orbit. The predicate {\tt expo m z t} asserts the {\em existence of a path} in an {\tt f}-orbit of {\tt m} from a dart {\tt z} to another {\tt t}, which is proven to be a {\em decidable equivalence}. 
Note that most of the properties are obtained by {\em noetherian induction} on the length of iterated sequences of {\tt f}-successors, bounded by the period.

Appropriate modules, called {\tt MA0}, {\tt MA1} and {\tt MF}, are written to instantiate for {\tt (cA m zero)}, {\tt (cA m one)} and {\tt (cF m)} definitions and properties of {\tt f}. So, a generic definition or property in {\tt Mf(M)} has to be prefixed by the module name to be concretely applied. For instance, {\tt MF.expo m z t} is the existence of a path from {\tt z} to {\tt t} in a face. In the following, {\tt MF.expo} is abbreviated into {\tt expf}. For instance, in Fig. \ref{fig:hmap2}, {\tt expf m 1 5 = True, expf m 5 3 = False}. Finally, a binary relation {\tt eqc} stating that two darts belong to the same component is easily defined by induction. For instance, in Fig. \ref{fig:hmap2}, we have {\tt eqc m 1 5 = True, eqc m 1 13 = False}. We quickly prove that {\tt (eqc m)} is a {\em decidable equivalence}. 

\vskip-.3cm
\subsection{Characteristics, Genus Theorem and Euler Formula}
\label{Ch}
We now count cells and components of a hypermap using the Coq library module {\tt ZArith} containing all the features of {\tt Z}, the integer ring, including tools to solve linear systems in Presburger's arithmetics. The numbers {\tt nd}, {\tt ne}, {\tt nv}, {\tt nf} and {\tt nc} of darts, edges, vertices, faces and components are easily defined by induction. Euler's characteristic {\tt ec} and {\tt genus} derive. The Genus Theorem and the Euler Formula (for any number {\tt (nc m)} of components) are obtained as corollaries of the fact that {\tt ec} is even and satisfies {\tt 2 * (nc m) >= (ec m)} \cite{duf07b}. Remark that {\tt ->} denotes a functional type in {\tt Set} as well as an implication in {\tt Prop}:
\begin{verbatimtab}
  Definition ec(m:fmap): Z:= nv m + ne m + nf m - nd m.
  Definition genus(m:fmap): Z:= (nc m) - (ec m)/2.
  Definition planar(m:fmap): Prop:= genus m = 0.
  Theorem Genus_Theorem: forall m:fmap, inv_hmap m -> genus m >= 0.
  Theorem Euler_Formula: forall m:fmap, inv_hmap m -> planar m -> 
       ec m / 2 = nc m.
\end{verbatimtab}

\section{Planarity and connectivity criteria}
\label{PC}
A consequence of the previous theorems is a completely constructive {\em criterion of planarity}, when one correctly {\em links} with {\tt L} at dimensions $0$ or $1$, {\em e.g.} for $0$:

\begin{figure}
\begin{center}
\includegraphics*[scale =.60]{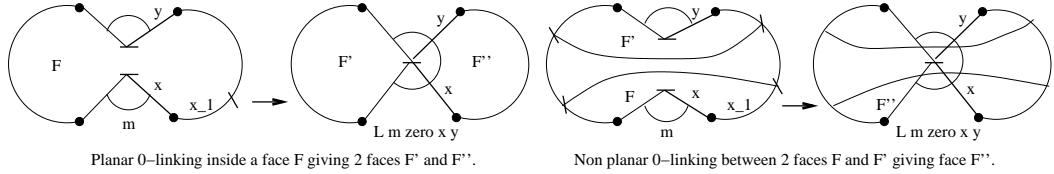}
\end{center}
\caption{Linking at dimension $0$.}
\label{fig:planarity}
\end{figure}

\begin{verbatimtab}
  Theorem planarity_crit_0: forall (m:fmap)(x y:dart), 
    inv_hmap m -> prec_L m zero x y -> (planar (L m zero x y) <->
      (planar m /\ (~ eqc m x y \/ expf m (cA_1 m one x) y))).
\end{verbatimtab}
So, the planarity of {\tt m} is preserved for {\tt (L m zero x y)} {\em iff} one of the following two conditions holds: (1) {\tt x} and {\tt y} are not in the same component of {\tt m}; (2) {\tt x}$\_${\tt 1 = (cA}$\_${\tt 1 m one x)} and {\tt y} are in the same face of {\tt m}, {\em i.e.} the linking operates {\em inside the face} containing {\tt y}. Fig. \ref{fig:planarity} illustrates $0$-linking inside a face, giving two new faces, and between two (connected) faces, giving a new face, thus destroying planarity. Finally, after a long development, we prove the expected {\em planarity criterion}, when breaking a link with {\tt B}, at any dimension, {\em e.g.} for $0$:
\begin{verbatimtab}
  Lemma planarity_crit_B0: forall (m:fmap)(x:dart), inv_hmap m -> 
    succ m zero x -> let m0 := B m zero x in let y := A m zero x  in
 (planar m <-> (planar m0 /\ (~ eqc m0 x y \/ expf m0 (cA_1 m0 one x) y))).
\end{verbatimtab}
Such a lemma is easy to write/understand as a {\em mirror form} of the  $0$-linking criterion, but it is much more difficult to obtain. It would be fruitful to relate these constructive/destructive criteria with the static one of G. Gonthier \cite{gon}. Finally, some useful results quickly characterize the effect of a link break on the {\em connectivity} of a planar hypermap. For instance, when $0$-breaking {\tt x}, a disconnection occurs {\em iff} {\tt expf m y x0}:
\begin{verbatimtab}
  Lemma disconnect_planar_criterion_B0:forall (m:fmap)(x:dart),
    inv_hmap m -> planar m -> succ m zero x ->
      let y := A m zero x in let x0 := bottom m zero x in
        (expf m y x0 <-> ~eqc (B m zero x) x y).
\end{verbatimtab}
\section{Rings of faces}
\label{FR}
\vskip-.3cm
\subsection{Coding a double-links and identifying a face}
\begin{figure}
\begin{center}
\includegraphics*[scale =.60]{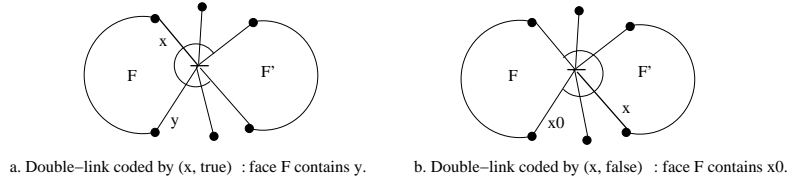}
\end{center}
\caption{Coding a double-link and identifying a face.}
\label{fig:identface}
\end{figure}

Since an edge is always open in our specification, when doing the backward break of a unique $0$-link from {\tt y} or {\tt y'}, we in fact realize a double-link break, as in Definition \ref{BaR}. 
So, we choose to identify a double-link by the unique dart, we called {\tt x}, where the $0$-link to be broken begins. 
In fact, with respect to the face {\tt F} {\em on the left of} the double-link in the ring, there are two cases, depending on the position of {\tt x} and its forward $0$-link, as shown in Fig. \ref{fig:identface} (a) and (b). We decided to distinguish them by a Boolean {\tt b}. Then, a double-link is coded by a pair {\tt(x, b)}.
So, we implicitely identify each ring face {\tt F} by the double-link coding on its right in the ring. In Fig. \ref{fig:identface} (a), face {\tt F} is identified by {\tt(x, true)} and contains {\tt y:= A m zero x}, whereas in Fig. \ref{fig:identface} (b), face {\tt F} is identified by {\tt(x, false)} and contains {\tt x0:= bottom m zero x}.
These modeling choices considerably simplify the problems. Indeed, in closed orbits, a true double-link break would entail $2$ applications of {\tt B} followed by $2$ applications of {\tt L}, and would be much more complicated to deal with in proofs.

\vskip-.3cm
\subsection{Modeling a ring of faces}
First, we inductively define linear lists of pairs of booleans and darts, with the two classical constructors {\tt lam} and {\tt cons}, and usual observers and destructors, which we do not give, because their effect is directly comprehensible:
\begin{verbatimtab}
  Inductive list:Set := lam: list | cons: dart*bool -> list -> list.
\end{verbatimtab}
Such a list is composed of couples {\tt (x, b)}, each identifying a face {\tt F}: if {\tt b} is {\tt true}, {\tt F} is represented by {\tt y:= A m zero x}, otherwise by {\tt x0:= bottom m zero x} (Fig. \ref{fig:identface}). In the following, {\tt Bl m l} breaks all the $0$-links starting from the darts of list {\tt l} in a hypermap {\tt m}. Now, we have to model the conditions required for list {\tt l} to be a ring of hypermap {\tt m}. Translating Definition \ref{RoF}, we have four conditions, called {\tt pre}$\_${\tt ringk m l}, for {\tt k} $=0, \ldots, 3$, which we explain in the following sections. Finally, a predicate {\tt ring} is defined by:

\begin{verbatimtab}
  Definition ring(m:fmap)(l:list):Prop:= ~emptyl l /\
    pre_ring0 m l /\ pre_ring1 m l /\ pre_ring2 m l /\ pre_ring3 m l.
\end{verbatimtab}

\vskip-.3cm
\subsection{Ring Condition (0): unicity}
The predicate {\tt distinct}$\_${\tt edge}$\_${\tt list m x l0} saying that the edges of {\tt l0} are distinct in {\tt m} from a given edge of {\tt x}, {\tt pre}$\_${\tt ring0 m l} is defined recursively on {\tt l} to impose that all edges in {\tt l} are distinct: Condition (0) of Definition \ref{RoF}. It also imposes that each dart in {\tt l} has a $0$-successor, in order to have well defined links, which is implicit in the mathematical definition, but not in our specification whith open orbits. 

\begin{verbatimtab}
  Fixpoint pre_ring0(m:fmap)(l:list){struct l}:Prop:=
   match l with 
     lam => True | cons (x,_) l0 => 
       pre_ring0 m l0 /\ distinct_edge_list m x l0 /\ succ m zero x
   end.
\end{verbatimtab}

\vskip-.3cm
\subsection{Ring Condition (1): continuity}
Then, we define adjacency between two faces identified
by {\tt xb = (x, b)} and {\tt xb' = (x', b')}, along the link corresponding to {\tt xb}:

\begin{verbatimtab}
  Definition adjacent_faces(m:fmap)(xb xb':dart*bool):=
   match xb with (x,b) => match xb' with (x',b') =>
   let y := A m zero x in let y':= A m zero x' in
   let x0 := bottom m zero x in let x'0:= bottom m zero x' in
     if eq_bool_dec b true 
     then if eq_bool_dec b' true then expf m x0 y' else expf m x0 x'0
     else if eq_bool_dec b' true then expf m y y' else expf m y x'0
   end end.
\end{verbatimtab}
This definition is illustrated in Fig. \ref{fig:adjacency}
for the four possible cases of double-link codings.
So, the predicate {\tt pre}$\_${\tt ring1 m l} recursively specifies that two successive faces in {\tt l} are adjacent: Condition (1) in Definition \ref{RoF}:

\begin{figure}
\begin{center}
\includegraphics*[scale =.60]{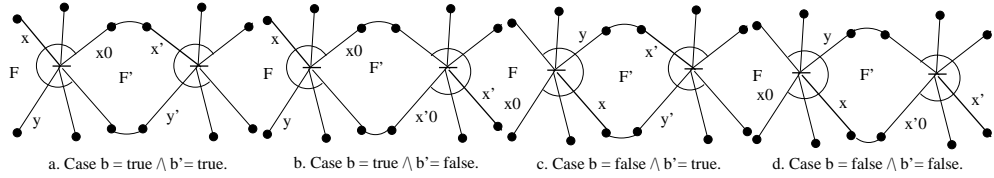}
\end{center}
\caption{Four cases of face adjacency.}
\label{fig:adjacency}
\end{figure}

\begin{verbatimtab}
  Fixpoint pre_ring1(m:fmap)(l:list){struct l}:Prop:=
   match l with 
    lam => True | cons xb l0 => pre_ring1 m l0 /\ 
      match l0 with lam => True | cons xb' l' => adjacent_faces m xb xb' end   
   end.
\end{verbatimtab}

\vskip-.3cm
\subsection{Ring Condition (2): circularity, or closure}
The predicate {\tt pre}$\_${\tt ring2 m l} specifies that the last and first faces in {\tt l} are adjacent: Condition (2) of circularity in Definition \ref{RoF}:

\begin{verbatimtab}
  Definition pre_ring2(m:fmap)(l:list):Prop:=
   match l with 
      lam => True | cons xb l0 =>
       match xb with (x,b) => let y := A m zero x in match l0 with 
         lam => let x0 := bottom m zero x in expf m y x0 
       | cons _ l' => let xb':= last l0 in adjacent_faces m xb' xb
       end end 
   end.
\end{verbatimtab}

\vskip-.3cm
\subsection{Ring Condition (3): simplicity}
The predicate specifiying that the faces of {\tt m} identified by {\tt xb} and {\tt xb'} are distinct is easy to write by cases on the Booleans in {\tt xb} and {\tt xb'}. The predicate {\tt distinct}$\_${\tt face}$\_${\tt list m xb l0} expressing that the face identified by {\tt xb} is distinct from all faces of list {\tt l0} entails. Then, the predicate {\tt pre}$\_${\tt ring3 m l} says that all faces of {\tt l} are distinct: Condition (3) in Definition \ref{RoF}: 

\begin{verbatimtab}
  Fixpoint pre_ring3(m:fmap)(l:list){struct l}:Prop:=
   match l with 
    lam => True | cons xb l0 => pre_ring3 m l0 /\ distinct_face_list m xb l0
   end.
\end{verbatimtab}

\section{Discrete Jordan Curve Theorem}
\label{JCT}
The general principle of the JCT proof for a hypermap {\tt m} and a ring {\tt l} is a structural induction on {\tt l}. 
The case where {\tt l} is empty is immediatly excluded because {\tt l} is not a ring by definition. Thus the true first case is when {\tt l} is {\em reduced to one element}, {\em i.e.} is of the form {\tt cons (x, b) lam}. Then, we prove the following lemma as a direct consequence of the  planarity criterion {\tt planarity}$\_${\tt crit}$\_${\tt B0} and the criterion {\tt face}$\_${\tt cut}$\_${\tt join}$\_${\tt criterion}$\_${\tt B0}:

\begin{verbatimtab}
  Lemma Jordan1:forall(m:fmap)(x:dart)(b:bool), inv_hmap m -> planar m -> 
     let l:= cons (x,b) lam in ring m l -> nc (Bl m l) = nc m + 1.
\end{verbatimtab}

When a ring {\tt l1} contains {\em at least two elements}, we prove that the condition \verb\~\{\tt expf m y x0} must hold with the first element {\tt (x,b)} of {\tt l1} (in fact, conditions $(1)$ and $(3)$ are enough):

\begin{verbatimtab}
  Lemma ring1_ring3_connect: 
   forall(m:fmap)(x x':dart)(b b':bool)(l:list), inv_hmap m ->
   let l1:= cons (x,b) (cons (x',b') l) in
   let y:=A m zero x in let x0:= bottom m zero x in 
     planar m -> pre_ring1 m l1 -> pre_ring3 m l1 -> ~expf m y x0.
\end{verbatimtab}
In this case, thanks to {\tt disconnect}$\_${\tt planar}$\_${\tt criterion}$\_${\tt B0} (Section \ref{PC}), the lemma entails that the break of the first ring link does never disconnect the hypermap. Then, after examining the behavior of {\tt pre}$\_${\tt ringk}, for {\tt k} $=0, \ldots, 3$, we are able to prove the following lemma which states that the four ring properties are preserved after the first break in {\tt l}:

\begin{verbatimtab}
  Lemma pre_ring_B: forall(m:fmap)(l:list), inv_hmap m -> planar m -> 
   let x := fst (first l) in let y := A m zero x in 
   let x0 := bottom m zero x in let m1 := B m zero x in 
 ~expf m y x0 -> ring m l -> (pre_ring0 m1 (tail l) /\ pre_ring1 m1 (tail l) 
      /\ pre_ring2 m1 (tail l) /\ pre_ring3 m1 (tail l)).
\end{verbatimtab}
The most difficult is to prove the part of the result concerning {\tt pre}$\_${\tt ringk}, for {\tt k} $=0, \ldots, 3$. The four proofs are led by induction on {\tt l} in separate lemmas. For {\tt pre}$\_${\tt ring0}, the proof is rather simple. But, for the other three, the core is a long reasoning where $2$, $3$ or $4$ links are involved in input. Since each link contains a Boolean, sometimes appearing also in output, until $2^4=16$ cases are to be considered to combine the Boolean values. 

Finally, from {\tt Jordan1} and {\tt pre}$\_${\tt ring}$\_${\tt B} above, we have the expected result by a quick reasoning by induction on {\tt l}, where links are broken one by one from the first:

\begin{verbatimtab}
  Theorem Jordan: forall(l:list)(m:fmap),
    inv_hmap m -> planar m -> ring m l -> nc (Bl m l) = nc m + 1.
\end{verbatimtab}
It is clear that, provided a mathematical hypermap $M$ and a mathematical ring $R$ conform to Definitions \ref{HD} and \ref{RoF}, we can always describe them as terms of our specification framework in order to apply our JCT. Conversely, given a hypermap term, some mathematical rings cannot directly be written as terms. 
To do it, our ring description and our JCT proof have to be slightly extended. However, that is not necessary for the {\em combinatorial maps} (where $\alpha_0$ is an involution) terms, for which our ring specification and our JCT formalization are {\em complete}. This is more than enough to affirm the value of our results.

\section{Conclusion}
\label{Cl}
We have presented a new discrete statement of the JCT based on hypermaps and rings, and a formalized proof assisted by the Coq system. Our hypermap modeling with {\em open orbits} simplifies and precises most of known facts. 
It also allows to obtain some new results, particularly about hypermap construction/destruction,
connection/disconnection and planarity. This work involves a substantial framework of hypermap specification, which is built {\em from scratch}, {\em i.e.} exempt from any proper axiom. It is basically the same as the one we have designed to develop geometric modelers via algebraic specifications \cite{ber:duf}. So, we know how to efficiently implement all the notions we formally deal with. 

The Coq system turned out to be a precious auxiliary to guide and check all the process of specification and proof.
The preexistent framework of hypermap specification represents about 15,000 lines of Coq, and the JCT development about 5,000 lines, including about 25 new definitions, and 400 lemmas and theorems. Note that all results about the dimension 0 were actually proven, but some planarity properties about dimension 1, which are perfectly symmetrical, have just been admitted. However, the JCT formal proof is complete. 

So, we have a solid foundation to tackle any topological problem involving orientable surface subdivisions. Extensions are in 2D or 3D computational geometry and geometric modeling by introducing embeddings \cite{duf:pui, ber:duf}, and computer imagery by dealing with pixels \cite{duf07a} or voxels.


\vskip-.3cm
\begin{thebibliography}{99}

\bibitem{bau:nip}
Bauer, G., Nipkow, T.:
The 5 Colour Theorem in Isabelle/Isar.
In \textit{Theorem Proving in HOL Conf.} (2002).
LNCS {\bfseries 2410}, Springer-Verlag, 67--82.

\bibitem{ber:duf}
Bertrand, Y., Dufourd, J.-F.: Algebraic specification of a 3D-modeler
based on hypermaps.  \textit{Graphical Models and Image Processing}
{\bfseries 56}:1 (1994), 29--60.

\bibitem{Coq}
The Coq Team Development-LogiCal Project:
The Coq Proof Assistant Reference Manual - Version 8.1, INRIA, France
(2007). \url{http://coq.inria.fr/doc/main.html}.

\bibitem{cor}
Cori, R.:
Un Code pour les Graphes Planaires et ses Applications.
\textit{Ast\'erisque} {\bfseries 27} (1970), Soci\'et\'e Math. de France.

\bibitem{deh:duf2}
Dehlinger, C., Dufourd, J.-F.:
Formalizing the trading theorem in Coq.
\textit{Theoretical Computer Science}  {\bfseries 323} (2004), 399--442.

\bibitem{duf:pui}
Dufourd, J.-F., Puitg, F.:
Fonctional specification and prototyping with combinatorial oriented maps.
\textit{Comp. Geometry - Th. and Appl.} {\bfseries 16} (2000), 129--156.

\bibitem{duf07a}
Dufourd, J.-F.:
Design and certification of a new optimal segmentation program with hypermaps.
\textit{Pattern Recognition} {\bfseries 40} (2007), 2974--2993.

\bibitem{duf07b}
Dufourd, J.-F.: A hypermap framework for computer-aided proofs in
surface subdivisions: Genus theorem and Euler's formula.  In:
\textit{22nd ACM SAC} (2007), 757--761.

\bibitem{fra95}
Fran\c{c}on, J.:
Discrete Combinatorial Surfaces.
CVGIP : Graphical Models and Image Processing {\bfseries 57}:1, (1995), 20--26.

\bibitem{gon}
Gonthier, G.:
A computer-checked proof of the Four Colour Theorem.  Microsoft
Research, Cambridge, \url{http://coq.inria.fr/doc/main.html} (2005), 57
pages.

\bibitem{gri}
Griffiths, H.:
\textit{Surfaces}.
Cambridge University Press (1981).

\bibitem{hal}
Hales, T.: A verified proof of the Jordan curve theorem.
Seminar Talk. Dep. of Math., University of Toronto (2005),
\url{http://www.math.pitt.edu/~thales}.

\bibitem{kor}
Kornilowicz A.: Jordan Curve Theorem.
In: \textit{Formalized Mathematics} {\bfseries 13}:4 (2005), Univ. of Bialystock, 481--491.

\bibitem{ros79}
Rosenfeld, A.:
Picture Languages - Formal Models for Picture Recognition.
In: \textit{Comp. Science and Appl. Math.} {\bf series}.
Academic Press, New-York (1979).

\bibitem{tut1}
Tutte, W.T.:
Combinatorial oriented maps.
\textit{Can. J. Math.}, {\bfseries XXXI}:5 (1979), 986--1004.
\end{thebibliography}
\end{document}